\documentclass[default]{sn-jnl}% Default
\jyear{2021}%

\theoremstyle{thmstyleone}%
\theoremstyle{thmstyletwo}%

\theoremstyle{thmstylethree}%

\begin{document}

\title[Star-forming dwarfs]{Chemical and stellar properties of star-forming dwarf galaxies}


\author*[]{\fnm{Francesca} \sur{Annibali}}\email{francesca.annibali@inaf.it}

\author*[]{\fnm{Monica} \sur{Tosi}}\email{monica.tosi@inaf.it}

\affil{\orgdiv{Osservatorio di Astrofisica e Scienza dello Spazio}, \orgname{INAF}, \orgaddress{\street{Via Gobetti 93/3}, \city{Bologna}, \postcode{40129}, \country{Italy}
}}


\abstract{Dwarf galaxies are the least massive, most abundant, and most widely distributed type of galaxies. Hence, they are key to testing theories of galaxy and Universe evolution. Dwarf galaxies sufficiently close to have their gas and stellar components studied in detail are of particular interest, because their properties and their evolution can be inferred with accuracy.
 This Review summarizes what is known of the stellar and chemical properties of star-forming dwarf galaxies closer than $\sim$20 Mpc. Given their low metallicity, high gas content and ongoing star formation, these objects are supposed to resemble the first galaxies that formed at the earliest epochs, and may thus represent a window on the distant, early Universe.
We describe the major results obtained in the past decade on the star formation histories, chemical abundances, galaxy formation and evolution of star-forming dwarfs, and the uncertainties still affecting these results.}




\maketitle

\section{Introduction}\label{sec1}
 
Dwarf galaxies are not only the most numerous type of observed galaxies, but also pivotal systems in the evolution of the Universe. The most popular cosmological scenario, the $\Lambda$ Cold Dark Matter  ($\Lambda$CDM) paradigm  \cite{White78,Peebles82}, predicts that galaxies form mainly through successive mergers or accretions of smaller systems. In this framework, dwarf galaxies are the first systems that collapse, form stars, pollute the surrounding intergalactic medium, leaking the ionizing photons emitted by their stars and ejecting the chemical elements produced by stellar nucleosynthesis, and then provide the basic building blocks that form larger systems like the Milky Way.  These properties make dwarf galaxies key ingredients in the processes that led to the so-called Epoch of Re-ionization in the Universe, to its progressive chemical enrichment, and to the formation and evolution of galaxies in general (see e.g. \citep{Tht09,Choi20} and references therein). As they are the seeds that lead to the formation of more massive galaxies, understanding the formation and evolution properties of dwarf galaxies is crucial to understand also massive ones.

Astronomers generally consider a dwarf any galaxy with baryonic (gas plus stars) mass not higher than about 10$^9$ M$_{\odot}$. Dwarf galaxies are usually classified on the basis of their morphology (late-type, such as irregular and blue compact dwarfs, and early-type, such as dwarf spheroidals, dwarf ellipticals, ultra faint dwarfs, etc.), or of their recent star formation (SF) activity (star-forming dwarfs vs quenched dwarfs). It has been suggested that quenched dwarfs (i.e., without gas and no evidence of recent SF) may be the result of the evolution of star-forming dwarf galaxies  (plenty of gas and currently forming stars) that become gas free because earlier SF consumed it all and/or because of environmental stripping  (e.g., \cite{Tht09}).
 In the smallest dark matter haloes re-ionization could have also suppressed gas cooling and quenched star formation (e.g., \cite{Efstathiou92}).  Detailed star formation histories suggest that quenched dwarfs had intense SF at early epochs, whilst current star-forming dwarf galaxies are those that started their SF activity rather slowly and weakly \cite{Gallart15}. 

Here we review the main stellar and chemical properties of star-forming dwarf galaxies (SFDs), i.e.,  those that, thanks to a conspicuous gas reservoir (their gas/baryonic mass fraction is around 50$\%$ or more), are still able to form stars. SFDs have a dark matter halo roughly ten times more massive than the baryonic component, and are found in any kind of environment, from the inner to the outer regions of galaxy clusters and groups to immense voids. Vice versa, early-type dwarfs are concentrated in galaxy groups and close to bigger galaxies. Because of their low mass, and their current high star formation rate, high gas content, and low metallicity, SFDs are the objects observable in the local Universe that most closely resemble the first galaxies that formed in the early Universe. Hence, studying these systems, scientists  can learn something about the first galaxies and how the early Universe evolved. 

\section{Star formation histories}\label{sec2}

Star formation is one of the key ingredients in galaxy evolution: it rules the stellar populations hosted in a region, its chemical enrichment, its capability to inject energy or pollution in the surrounding medium. How the star formation varied with time within a galaxy, i.e. its star formation history (SFH), is therefore crucial to understand its evolution. 

The main questions that SFH studies of dwarf galaxies need to address are: How many dwarfs (if any) had SF sufficiently intense at the earliest epochs to ionise the surrounding Universe? Are there SFDs that have started forming their stars only recently? Is the SF regime continuous or not? Is there a substantial fraction of SFDs suffering very strong SF bursts (i.e. with rate more than an order of magnitude higher than the average rate)? What is the spatial distribution of the SF activity?

Until about 20 years ago, astronomers (see, e.g., \cite{mc83,marconi94}) thought that  the majority of SFDs are galaxies experiencing strong bursts of SF separated by long quiescent phases, but the wealth of studies 
triggered by the advent of the Hubble Space Telescope (HST) have shown that this is not the case (e.g., \cite{Tht09} and references therein). 

Thanks to the high-spatial resolution of HST photometry, it has been possible to resolve individual stars and draw their colour-magnitude diagrams (CMDs) in galaxies as distant as about 20 Mpc. The so-called synthetic CMD method, i.e., the comparison of the observational CMD with simulated ones, based on state-of-the-art stellar evolution models and built taking into account all the intrinsic (e.g., distance, reddening, initial mass function, stellar multiplicity) and observational uncertainties (e.g., photometric errors and incompleteness), is in fact the most robust method (e.g. \cite{Tosi91,Dolphin02,Gallart05,Ct10,Cignoni15}) to derive the SFH of a galactic region.
The SFHs of dwarf galaxies as derived from the CMDs of their resolved stellar populations were reviewed about ten years ago by \cite{Tht09}, and we refer to that paper for the results published until 2009.
 In the last decade, several interesting studies have been devoted to deriving the SFHs of SFDs located  inside and outside the Local Group to fully address the questions that at the time had only received partial or uncertain answers.
 
 % Figure 1
\begin{figure*}
\centering
\includegraphics[width=0.99\linewidth]{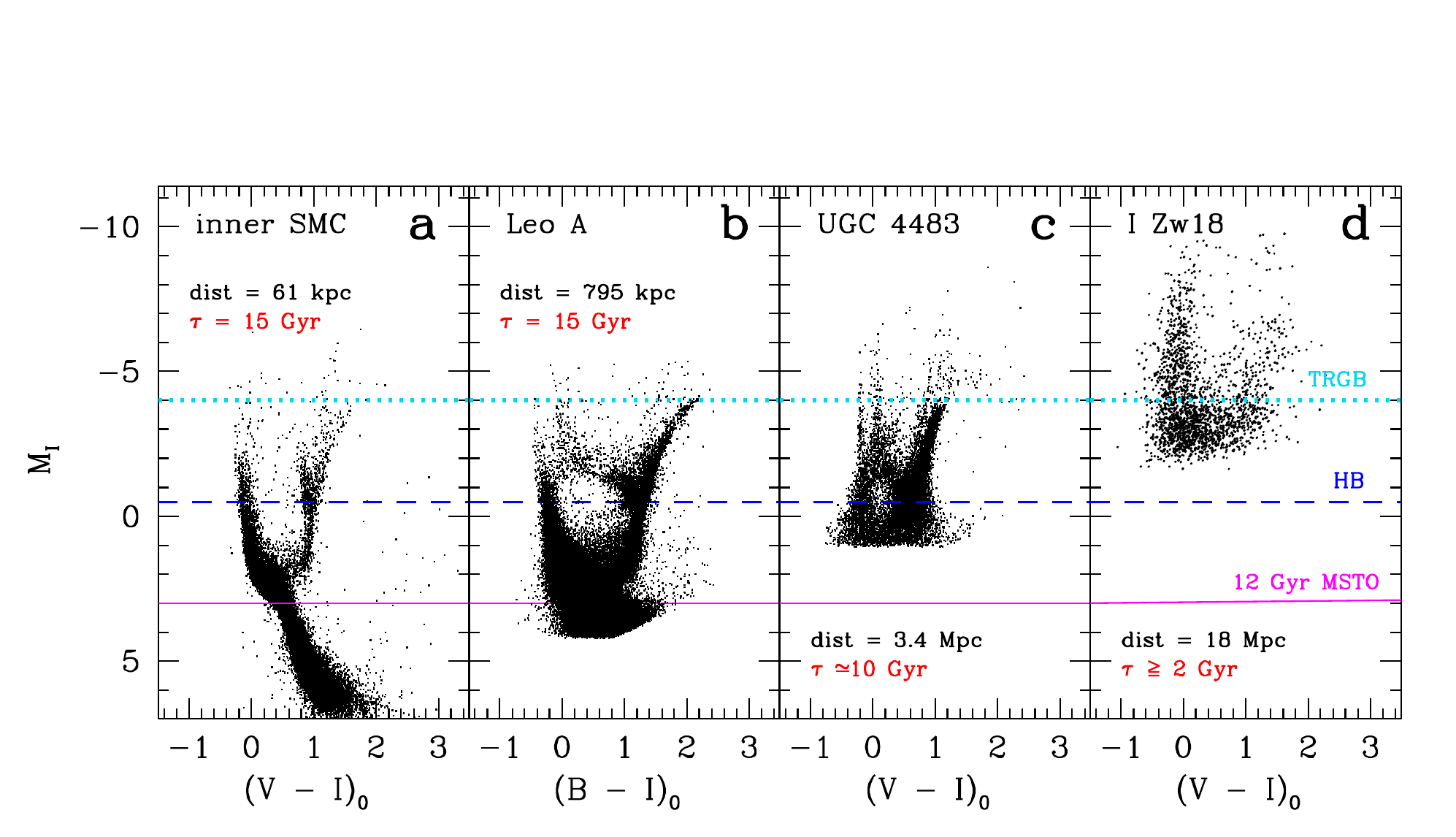}
\caption{The effect of distance on observational CMDs.  Absolute magnitude vs intrinsic colour of SFDs  at increasing distances are plotted from panel a to d. The horizontal lines show the magnitude level of the evolutionary phases used as old age indicators (see text for details): RGB tip in dotted cyan, Horizontal Branch in dashed blue, 12 Gyr main-sequence turnoff in solid magenta.  The lookback time corresponds to the age of the oldest stars identified in the CMD. Galaxy name, distance and lookback time are labelled in each panel. The dwarfs plotted in panels  a and b are within the  Local Group, and their oldest turnoff is visible. Farther out, at 3.4 Mpc, HST allows to (barely) reach the Horizontal Branch, hence stars 10 Gyr old (panel c); at 18 Mpc, the oldest stars resolvable with accuracy are on the RGB, with age $\geq$1-2 Gyr  (panel d).  The photometric uncertainty increases with increasing distance: for instance, at the RGB tip it ranges from 0.01 mag in panel a to 0.20 mag in panel d.
}
\label{cmdabs}
\end{figure*}

To derive the SFH over the whole lifetime of a galaxy,  one needs to measure luminosity and colour of resolved stars of all ages, convolve them with stellar evolution and assumptions on the initial mass function, and estimate how many stars have formed at any epoch. The inevitable observational limitation is that the farther the system, the intrinsically brighter  (i.e., more massive or more evolved) the stars that one can resolve.  This is apparent in Fig.\ref{cmdabs}, where we show the CMDs of four SFDs located at increasing distances and observed by HST with similar setups. The distance increase makes it progressively more difficult to detect very old stars, because, if they are still alive, they have small mass and are therefore rather faint. 
In practice, only in Local Group galaxies can the telescopes currently in operation allow to measure  old unevolved stars still on the main-sequence, the longest and most prominent stellar evolutionary phase, whose turnoff point provides the only reliable clock to age-date stars older than $\sim$10 Gyr outside our Galaxy \citep[e.g.][and references therein]{Chaboyer95}. 
%as old as the Universe, 
%{\bf i.e. 13.8 Gyr \citep[ref.][]{Planck2020}.}

In order of increasing intrinsic brightness and, hence, of increasing visibility distance limit, the other evolutionary phases that are reliable clocks for old populations in a system are the Horizontal Branch,  corresponding to stars at least 10 Gyr old, the Red Clump for stars 3-4 Gyr old, and the Red Giant Branch (RGB) for stars at least 1-2 Gyr old. % {\bf If these evolutionary sequences are visible in the CMD of a system, the system contains stars as old as the age typical of those evolutionary phases, and these stars are resolved with good accuracy by the available photometry. }
The presence of these sequences in the CMD indicates that the system contains stars  at least as old as the age typical of those evolutionary phases. 
As of today,  with the photometric depth reachable with the most powerful existing instruments, the Horizontal Branch can be detected in galaxies closer than 3-4 Mpc, the Red Clump in galaxies closer than 5-6 Mpc, the RGB in galaxies up to 18-20 Mpc  (see Fig.\ref{cmdabs}).

As a result, only for SFDs within the Local Group it is currently possible to robustly derive accurate SFHs back to the earliest epochs. This goal has been successfully met using mainly HST photometry, and nowadays almost all the dwarf galaxies in the Local Group have their SFH accurately derived from their CMDs (e.g., \cite{Skillman14,Gallart15,Albers19}, and references therein). All these studies have shown that SF occurred in  local SFDs since the earliest epochs, starting with a low rate maintained for several Gyrs and then increased significantly in the last few billion years. 
Where the spatial distribution of the SF was also analysed, it turned out to be currently rather patchy, but always higher in the inner than in the outer regions, with a clear positive age gradient (see, e.g. WLM \cite{Albers19}).  Positive age gradients are found also in nearby dwarf spheroidals and transition-type dwarfs (see, e.g. \cite{Hidalgo09,Hidalgo13}), and are expected on the basis of galaxy evolution simulations taking into account mergers and star formation feedback, such as FIRE (Feedback In Realistic Environments). They are predicted to be steeper when the bulk of SF occurs earlier \cite{Albers19,Graus19}. 

\begin{figure*}
\centering
\includegraphics[width=0.99\linewidth]{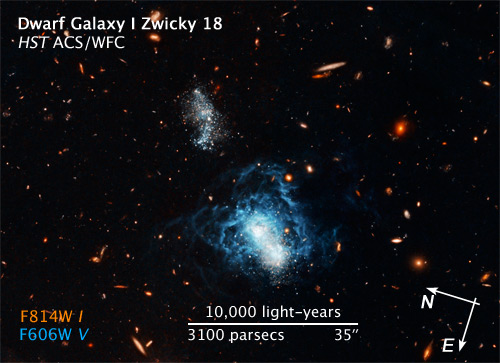}
\caption{ The sharp view from HST of SFDs.  As an example, we show the blue compact dwarf galaxy IZw18, and its secondary component to the north-west, as imaged by the Advanced Camera for Surveys on board HST in the F606W (blue) and F814W (red) filters (\cite{Aloisi07} and the Hubble Heritage. Credits NASA, ESA, and A. Aloisi).}
\label{izw18}
\end{figure*}

In spite of the higher accuracy achievable in the SFH derivation of the nearest galaxies, it is mandatory to face also the challenge of more distant targets, because the Local Group does not host all types of SFDs. In particular, it does not contain blue compact dwarfs, strong starbursts, and extremely metal poor ones. These are quite interesting sub-types of dwarfs,  more active and, often, more metal poor than average; hence, more similar to primeval galaxies.
%The enterprise of deriving the SFH of SFDs beyond 2 Mpc started in the mid '90s, after the first refurbishment of HST that restored its sharp view, and then proceeded at accelerating pace when the telescope was equipped with the ACS and, later, the WFC3 cameras. 
Among the SFDs located outside the Local Group whose HST-derived CMDs were used to infer the SFH we recall the 18 starbursts with distances up to 8 Mpc analysed by \cite{Mcquinn10}, NGC~1569, a strong starburst dwarf at $\sim$3 Mpc   \cite[ref.][]{Greggio98,Grocholski12},  NGC~1705, a blue compact dwarf at $\sim$5 Mpc  \cite[ref.][]{Annibali03,Annibali09}, and IZw18, the iconic prototype of extremely metal-poor blue compact dwarfs,  at 18 Mpc  \cite[ref.][]{Aloisi07,Annibali13}. The latter still holds the record of the most distant galaxy with SFH derived from the CMD (shown in panel d of Fig.\ref{cmdabs}) of its resolved stellar populations. The HST image of IZw18 is shown in Fig.\ref{izw18}.

% Figure 2 
% Figure 2

A great boost to this research field was triggered by the HST ACS Nearby Galaxy Survey Treasury (ANGST \cite{Dalcanton09}) that obtained uniform multi-colour photometry of the resolved stars in 69 galaxies residing in diverse environments within 4 Mpc.  From the CMDs  of the ANGST targets, \cite{Weisz11}  derived the SFHs of the 28 SFDs in that sample. They concluded that about two thirds of the studied SFDs are experiencing ongoing episodes of SF activity $\sim$3 times stronger than in the past, and that enhanced SF episodes last on average several hundreds Myr.
More recently, another HST Treasury program, LEGUS (Legacy ExtraGalactic Ultraviolet Survey, \cite{Calzetti15}) %{\bf  inferred} the CMD-derived SFHs of 
imaged in the UV star forming galaxies up to $\sim$12 Mpc. %Thanks to the UV sensitivity of the LEGUS data, {\bf that helps to better isolate the central He-burning sequences in the CMD, detailed temporal and spatial properties were derived for the star formation activity} occurred in the last 50--100 Myr in 24 SFDs \cite{Cignoni18,Cignoni19}.
 Thanks to the combination of deep UV and optical photometry, the star formation history occurred in the last  50--100 Myr  was derived with high temporal and spatial details in 24 SFDs \cite{Cignoni18,Cignoni19}.
 In none of these dwarfs the current SF rate was found to be higher than a few times the average rate of the recent epochs, thus confirming that extreme starbursts are quite rare events. 
This result confirms, on the basis of resolved stellar population studies, what was found by \cite{Lee09} on the basis of the integrated H${\alpha}$ emission of 300 star-forming galaxies within 11 Mpc: systems currently experiencing massive global bursts are just $\sim6\%$ of the SFD sample, and bursts are only responsible for about a quarter of the total star formation in the overall dwarf population.

Of special interest are extremely metal-poor dwarfs (XMPs, see Section 3.1.1), because their very low oxygen abundance (12+log(O/H)$\leq$7.35  \cite[ref.][]{Guseva15}, or less than about 5$\%$ of solar), low mass and high gas content make them most similar to primordial galaxies. Initially they were thought to be actually young systems experiencing now their first SF activity \citep[e.g.,][]{IT04}, but deep HST photometry reveals that all of them host stars as old as the reachable lookback time \cite{Aloisi07,Sacchi16,Sacchi21,Mcquinn15,Mcquinn20}, just like all the other galaxies studied so far. It is thus crucial to accurately derive their SFH back to the earliest epochs, and figure out how it may be compatible with such a low chemical enrichment (see next sections).
For this reason, in the last few years several groups have focused on the search of XMPs and on the derivation of their SFH. Among those that have been discovered within a distance of $\sim$20 Mpc, 
%the local volume {\bf (defined as the portion of Universe within about 20 Mpc from Earth)
hence within reach of resolved stellar population studies, we recall here DDO68, UGC4483, Leo~P, and Leoncino. 

DDO68 is one of the most intriguing XMPs \cite{Pustilnik05} and provides direct observational evidence of multiple galaxy merging occurring at very low galactic mass scales. Located at almost 13 Mpc, in the huge Lynx-Cancer void \cite{Pustilnik11}, thus supposed to be one of the most isolated galaxies, it immediately revealed a very distorted morphology, with a sort of cometary tail very suggestive of gravitational interactions  \cite{Cannon14,Tikhonov14,Sacchi16}. Deep  wide-field photometry obtained with the Large Binocular Telescope, combined with HST imaging, confirmed that DDO68 interacts with at least two smaller-mass companions \cite{Annibali16}. From the CMD of their resolved stars,  it turns out that the SF in the smallest satellite was likely quenched by the interaction \cite{Annibali19a} about 1 Gyr ago, while that in the  intermediate mass companion shows a significant enhancement in the last few hundreds Myr  \cite[ref.][]{Sacchi16}, suggesting that the accretion is ongoing or quite recent.

At a distance of $\sim$3 Mpc  \cite[ref.][]{Sacchi21}, UGC4483 is the closest XMP \citep[see][]{Skillman94} blue compact dwarf. Its SFH was first derived by \cite{Mcquinn10}, from a CMD reaching down to the RGB magnitude level, hence with a minimum lookback time of a few Gyrs. New and deeper photometry acquired with the most recent HST camera has lately allowed \cite{Sacchi21} to identify the Horizontal Branch in the CMD of its less crowded regions, as well as the presence of candidate RR~Lyraes (10 Gyr old pulsating stars). The CMD of UGC4483 is shown in  panel c of Fig.\ref{cmdabs}. The new SFH thus covers the entire Hubble time, and it is interesting to emphasise that this history is qualitatively similar to that found in the SFDs in the Local Group: an early slow start, at a steady low SF rate lasting several Gyrs, followed by a higher rate since  $\sim$1 Gyr ago. Equally similar is the spatial age gradient, with the younger stars mostly concentrated in the inner regions, and the older populations dominating the outer ones. The strongest SF peak is the one currently ongoing in the innermost regions, and none of the peaks exceeds the past average rate by more than a factor of 3. 
%Generally speaking there is no evidence that BCDs have much stronger SF bursts that dwarf irregulars. Actually there are cases  where the SFR of a dIrr (e.g. NGC1569 \cite{Greggio98}) is a factor of 100 higher than in average BCDs.

LeoP is an XMP \citep[see][]{Skillman13} with very low mass and luminosity, discovered by the Arecibo  Legacy  Fast  ALFA  blind  HI survey at a distance of $\sim$1.6 Mpc, in a fairly isolated region at the edge of the Group. Thanks to its proximity, the CMD resulting from follow-up HST photometry clearly shows the Horizontal Branch, confirmed also by the identification of RR Lyrae variables.  This allowed \cite{Mcquinn15} to robustly infer its SFH back to 10 Gyr ago, concluding that the SF started at the earliest epochs, then 
proceeded at a quite low rate for about 8-10 Gyr, and then increased to a higher, roughly constant rate in the last 4 Gyr, similarly to what is seen in more massive SFDs. \cite{Mcquinn15} suggest that LeoP has lost 95$\%$ of its produced oxygen, likely via stellar-feedback-driven galactic winds.

Leoncino is another XMP \citep[see][]{Hirschauer16} that was found by the Arecibo survey, and then followed up with HST photometry to resolve its stars and infer both its distance and its SFH. It turns out to be located at $\sim$12 Mpc, in a void \cite{Mcquinn20}. Quite interestingly, as in the case of DDO68,  although in a very low density environment, Leoncino shows signs of interaction with another galaxy in the same void, UGC5186, at a projected distance of only 46 kpc, possibly responsible for the recent SF activity.

To summarise, in the last decade astronomers have definitely assessed that: 1) SFDs have all started forming stars at the earliest epochs; 2) their SF is rather continuous, although with significant enhancements usually lasting from hundreds to thousands Myr; 3) SF bursts with rates higher than ten times the average rates are rare, at least at time resolutions of $\geq$200 Myr;  4) the spatial distribution of SF regions is patchy, with younger stars usually more numerous towards the central regions. It remains to: a) reliably quantify
what was the average star formation rate at the earliest epochs and whether it was sufficient to contribute to the Universe re-ionization; b) assess the role of re-ionization in inhibiting star formation in the smallest structures 
(M$_{\star}\lesssim10^6$ M$_{\odot}$); and c) quantify the impact of gravitational encounters on the SF of dwarf galaxies.

\section{Chemical properties}\label{sec3}

The present-day distribution of metals across dwarf galaxies results from a complex interplay between chemical enrichment, loss of metals, gas accretion and feedback. 

Studies of the chemical properties in SFDs are mostly based on optical and near-infrared spectroscopy of bright HII regions 
%\cite{PagelEdmunds81,Skillman89a,Skillman89b,Skillman94,Lee03,Lee04,VZ06,Berg12,Haurberg15,Jimmy15}
 \cite{PagelEdmunds81,Skillman89a,VZ06,Berg12,Haurberg13} since, even with the most advanced instrumentation on 8-10 m telescopes, metallicity measurements in individual (giant and super-giant) stars are only feasible at distances $\lesssim$1 Mpc \citep[e.g., ][]{Venn01,Kaufer04,Bresolin06,Kirby17,Hermosa20}.
Comparison of population synthesis models with multi-band photometry and stellar absorption lines measured  in the integrated-light spectra of galaxies and star clusters has sometimes been pursued to infer the stellar chemical content of SFDs \citep{Whitmore20,Watson21}, but the uncertainties provided by this approach are typically large 
($\sim$0.3-0.5 dex).  
%\citep{Venn01,Venn03,Kaufer04,Bresolin06,Taut07,Kirby12,Hosek14,Kirby17,Hermosa20}.
 As a consequence,  our view of the chemical properties in SFDs is highly biased toward the current gas composition around regions of star formation, while direct information on both the global ISM composition and the past chemical enrichment are rather scarce. 
 
Planetary nebulae (PNe), which are the end product of stars with mass $\lesssim$8 M$_{\odot}$, offer the possibility of inferring the local composition at relatively earlier epochs ($\gtrsim$100 Myr ago) and, thanks to their bright emission lines, have been used as chemical tracers in SFDs up to distances of a few Mpc. The combined analysis of HII regions and PNe in principle provides useful constraints to chemical evolution models, however, the uncertainty in the derivation of the PN's age \citep[up to several Gyr, see e.g.][]{Magrini09} and on the effects of environmental conditions make the interpretation far from unique. For instance, the similar oxygen abundances derived from the two types of emission objects in most analyzed SFDs have led \citep{Magrini05,Richer07,Magrini09,Annibali17} to suggest that the modest chemical enrichment could be due to recent  loss of metals through galactic winds, while the lower HII region metallicities compared to PNe in other SFDs have been interpreted by \cite{Pena07} as dilution through accretion of primordial gas or interaction with a metal poorer companion. Accretion and interaction events could help explain the extremely low metal content measured in some XMPs 
with relatively large masses, which are strong outliers in the mass-metallicity relation.

Equally important are studies of the neutral gas composition, since 
this phase is 
likely less affected by self-pollution of very recently produced metals \citep{Kunth86}.  HI metallicity studies in SFDs based on ultraviolet absorption lines have shown systematically lower element abundances in the neutral ISM compared to the ionized gas (e.g., \cite{Kunth94,Aloisi03,Cannon05,Bowen20} and references therein), although the results presented by \cite{Kunth94} were questioned by \cite{Pettini95}.

\subsection{The SFD mass-metallicity relation}

\begin{figure}
\centering
\includegraphics[width=9cm]{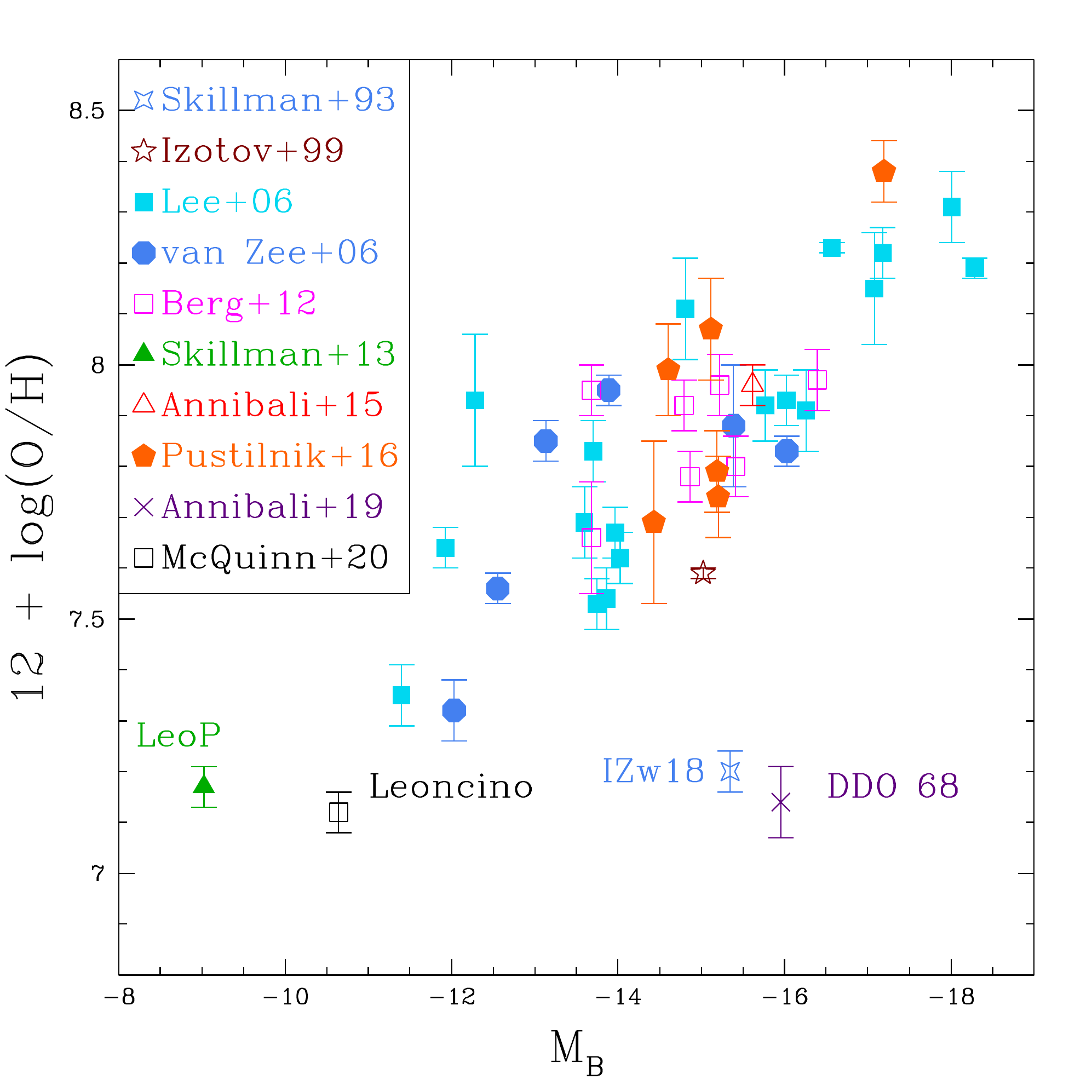}
\caption{ The luminosity-metallicity relation for SFDs. HII region oxygen abundances,  with their 1-$\sigma$ uncertainties, versus absolute B magnitudes for SFDs in the literature, restricted to galaxies with 
direct T$_e$ method measurements and robust distance determination from the RGB tip. The plotted O/H values are from
\citep{Skillman93,IT99,Lee06,VZ06,Berg12,Skillman13,Annibali15,Pustilnik16,Annibali19b,Mcquinn20}. The names of some famous XMPs are indicated in the plot.}
\label{mz}
\end{figure}

Among galaxy scaling relations, the mass-metallicity (M-Z) and luminosity-metallicity (L-Z) relations 
\citep[e.g., ][]{Lequeux79a,Skillman89a,Pilyugin01,Garnett02,Berg12,Hunt16,Curti20} bear crucial information on the SFH, initial mass function, gas outflow and inflow, and merging history of dwarf galaxies \cite{LSC06,Brooks07}. 
In order to reliably constrain both the slope and the intrinsic scatter of these relations, accurate (below 0.1 dex uncertainty) measurements of HII region chemical abundances  are needed; these can be obtained through the so-called 
direct-T$_{e}$ method which requires the detection of weak temperature-sensitive auroral lines (e.g.,[OIII]$\lambda$4363 and [NII]$\lambda$5755) and it is thus confined to bright HII regions in a limited number of systems.
 Fig.~\ref{mz} shows the L-Z relation for SFDs, based on a collection of direct-T$_{e}$ oxygen abundances derived from the literature. 
 
%Fig.3
% Figure 3 

As we will discuss in section~\ref{sec_chemev}, the metallicity decrease observed in SFDs with decreasing mass/luminosity is mainly explained as due to the combined effect of progressively less efficient star formation and increased susceptibility to metal loss  
through galactic winds in shallower gravitational potential wells. 
Also, void galaxies appear  to be slightly biased toward lower metallicities compared to systems in higher density  environments, possibly due to lack of interactions and mergers, and less-efficient star formation in very low density environments  \cite{Pustilnik16,Kniazev18}.

%The most accurate determinations of ionized gas abundances of several elements (He, N, O, Ne, Ar, and S) in dwarfs rely on the direct-Te method \cite{Dinerstein90} which requires the detection of weak temperature-sensitive auroral lines (e.g.,[OIII]$\lambda$4363 and [NII]$\lambda$5755) and it is thus confined to bright HII regions in a limited number of systems (). Therefore, oxygen abundance measurements often rely on strong-line calibrations \cite{Pettini04,pilyugin05,Yin07,pilyugin16}, carrying significant uncertainties and systematic effects (see \cite{MM19} for a review). 

%\subsubsection{M-Z and L-Z relations}
%Minimizing uncertainties in abundance measurements is crucial for an accurate characterization of observed dwarf galaxy scaling relations, such as the mass-metallicity (M-Z) and luminosity-metallicity (L-Z) relation \cite{Lequeux79,Skillman89a,Pilyugin01,Garnett02,Berg12,Haurberg15,Curti20}, where both the slope and the intrinsic scatter bear information on the star-formation history, initial mass function, gas outflow and inflow, and merging history \cite{LSC06,Brooks07,Sanders20}

\subsubsection{Extremely metal-poor galaxies}

Galaxies at the  lowest metallicity end of the relation displayed in Fig.~\ref{mz} are of particular interest as benchmarks to understand star formation processes occurring in the early Universe \citep[e.g., ][]{KO00,Skillman93,Pustilnik05,Izotov06a,Skillman13,Mcquinn20}. They are also the favoured targets for emission-line spectroscopy aimed at  inferring the primordial $^4$He abundance which, provided that it is derived with an accuracy better than one per cent, allows to constrain Big Bang Nucleosynthesis \cite{Yang84,Walker91,IT98,Izotov2007}.  In this respect, significant progress in reducing systematic uncertainties in the determination of  the $^4$He abundance in metal-poor galaxies has been 
achieved in recent years thanks to the addition of near-infrared helium lines 
\cite{Izotov2014}  and to the implementation of improved statistical methodologies for the data analysis \cite{Aver2015,Fernandez2019,Valerdi2019,Hsyu2020,Aver2021,Kurichin2021}.

Galaxies are defined as extremely metal-poor (XMP) if they have an oxygen abundance 12+log(O/H)$\leq$7.35  \cite[ref.][]{Guseva15}.
There is some confusion in the literature on the upper limit to the oxygen abundance of  this class of objects, due to a seminal paper by \cite{KO00}, who defined as  ``very metal poor" galaxies with less than 12+log(O/H)$\leq$7.65. Here we follow the more restrictive definition by \cite{Guseva15}. 
%\cite{Skillman93,Pustilnik05,Izotov06a,Skillman13,Hirschauer16,Hsyu17,Izotov18a,Kojima20,Mcquinn20}, 
XMPs are quite rare galaxies, but because of their interest they are currently being actively searched for by several groups \citep[e.g., ][]{Izotov12,Skillman13,Hirschauer16,Guseva2017,Hsyu2017,Yang2017,Izotov2018,Hsyu2018,Senchyna2019a,Senchyna2019b,Kojima2020,Mcquinn20,Pustilnik2020,Pustilnik2021}.

%A few objects in Fig.XXX display oxygen metallicities lower than 12 + log(O/H) $\sim$ 7.3 or $\sim$1/30 solar
 %and are typically referred as to extremely metal poor galaxies (XMPs)
%These systems resemble primeval galaxies, and are benchmarks both to study star formation processes in the early Universe, and to infer the primordial helium abundance (see section xxx).
Some XMPs fit the M-Z and L-Z relations (e.g., Leo P and Leoncino) and their properties can be explained with secular evolution characterized by inefficient star formation and metal loss via galactic winds at low galaxy masses. Other XMPs are instead strong outliers in both relations and have been suggested to have experienced accretion of metal poor gas not only from the intergalactic medium but also as a consequence of  interaction events  \cite{Ekta10,Mcquinn20}, a scenario often supported by the presence of disturbed HI morphology and kinematics \cite{Lelli12,Cannon14}. An iconic example of this situation is the XMP DDO~68 already discussed in Section~\ref{sec2},
whose extremely low HII regions metallicity of a few percent solar, incompatible with its stellar mass of $\sim10^8$ M$_{\odot}$, has been suggested to actually reflect the composition of its ten times smaller, accreted satellite rather than that of the main galaxy \citep{Tikhonov14,Pustilnik05}. This scenario has been recently confirmed by hydrodynamical simulations \cite{Pascale21}.

\subsection{Element abundance ratios}

Element abundance ratios provide constraints to models of galaxy evolution and chemical enrichment.
In particular, the ratio of the $\alpha$-elements over iron-peak elements, [$\alpha$/Fe], is known to 
be tightly linked to both the stellar initial mass function and  the star-formation timescale, because the $\alpha$-elements  
are mainly synthesized by massive stars and restored into the ISM within $\sim$50 Myr from the star formation onset,  
 while Fe, mostly provided by SNIa, is injected into the ISM on longer time-scales
\citep{Greggio83,Matteucci01}. Unfortunately, while optical and near-IR spectra contain bright emission lines of several 
$\alpha$-elements, such as O, Ne, Ar, and S, measurements of Fe require very deep observations because of the weakness of its lines in the optical; furthermore, Fe abundances carry significant uncertainties due to depletion into dust grains,  hampering  a meaningful characterization of $\alpha$/Fe ratios in HII regions.
Integrated-light spectroscopy of stellar absorption features in globular clusters provides an alternative pathway to the derivation of $\alpha$/Fe ratios, and this approach has indeed been pursued in a few nearby SFDs \citep{Strader03,Puzia08,Sharina10,Annibali18}. These studies indicate typically low  $\alpha$/Fe ratios, 
consistent with measurements in individual stars \citep{Venn01,Kaufer04}; such values are expected 
for galaxies that have formed stars at a low rate over a long period of time, as indeed is the case for SFDs (see Section 2), and where galactic outflows may have possibly contributed to the removal of the $\alpha$-elements. 

% Figure 4
% Figure 4 
\begin{figure}
\centering
\includegraphics[width=9cm]{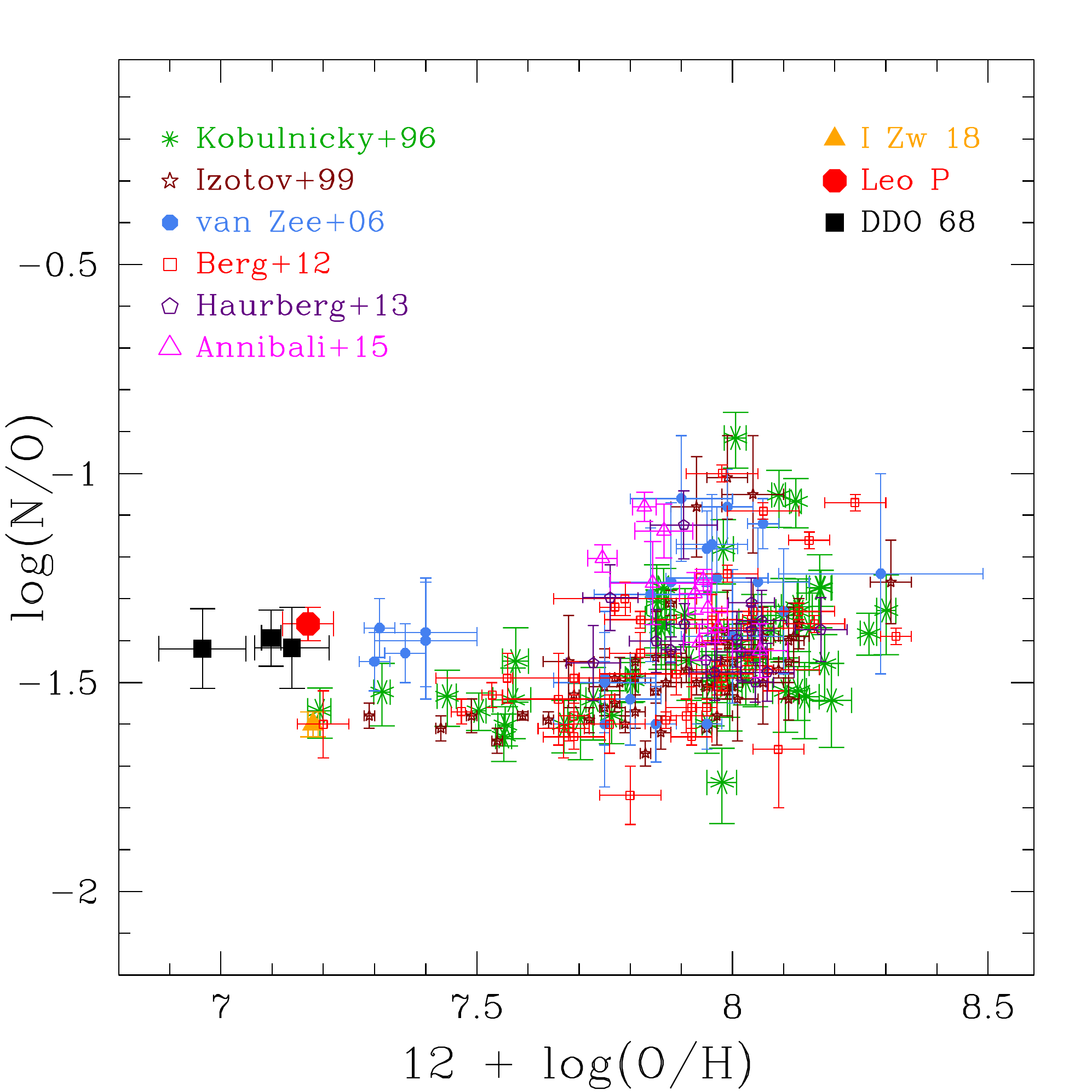}
\caption{ Element abundance ratios in SFDs. HII region N/O versus O abundances for SFDs in the literature, restricted to galaxies with 
direct T$_e$ method measurements.  Error bars refer to 1-$\sigma$ uncertainties. Plotted values are from 
\citep{Kobulnicky96,IT99,VZ06,Berg12,Haurberg13,Annibali15}. The XMPs galaxies IZw18 \citep{Skillman93,IT99}, 
Leo P\citep{Skillman13}, and DDO~68 \citep{Annibali19b} are indicated.}
\label{nsuo}
\end{figure}

In HII regions, the observed flat trends of  $\alpha$-element ratios as a function of oxygen abundance confirms that these elements are all synthesized by massive stars on similar timescales, thus their variation occurs in lockstep. 
On the other hand, nitrogen, whose behavior is presented in Fig.~\ref{nsuo}, is more complex.
While the observed growth of N/O with increasing O/H abundance 
in all galaxies with 12 + log(O/H) $\geq$ 7.5- 8.0  \citep[ref.][]{Pagel85,vanZee98} is explained with the presence of secondary nitrogen \cite{Tinsley80}
(i.e., produced using as catalyst the carbon already present in the proto-star cloud), 
the flat plateau at log(N/O)$\sim$-1.6 for 12 + log(O/H) $\leq$ 7.5 has been suggested to 
be  due to primary nitrogen produced by low-metallicity massive stars and released into the ISM on similar timescales 
as oxygen \cite{Timmes95,IT99}. 
A similar behaviour has been observed for carbon: a flat C/O trend (but with a large dispersion) 
 in metal poor dwarf galaxies followed by a steady C/O increase at 12 + log(O/H)$\geq$8  \cite[ref.][]{Berg2016,Berg2019}.
 % , suggesting  that 
%carbon may follow nitrogen in originating from
%primary production at low O/H and from  pseudo-secondary processes at higher metallicities  

%, a scenario supported by stellar evolution models with rotation (e.g.  Maeder \& Meynet (2001)).  
 Recent abundance measurements in XMPs present a new challenge, as some of them  exhibit  N/O ratios above the plateau \citep[e.g., ][]{VZ06,Vanzee06,Berg12,Izotov12}.
Firm conclusions for this unexpected behavior 
have not been reached yet.
Some authors \cite{Pilyugin92,Henry00} propose a scarce oxygen production from an inefficient current SF compared to an increased nitrogen enrichment from intermediate mass-stars generated during a stronger,  older burst.  Another possibility is N enrichment by Wolf Rayet stars close to  H II regions, as suggested by detailed studies of chemical spatial properties in SFDs  (see section~\ref{sec_spatial}). 
Also for carbon, determining the true nature and dispersion of the observed C/O trend at low O/H requires larger galaxy samples than those analyzed so far. This task is difficult to accomplish, because such measurements rest on deep UV spectroscopy \citep[e.g., ][]{Garnett1995,Berg2016}.

\subsection{Chemical spatial properties} \label{sec_spatial}

%The present-day distribution of metals across dwarf galaxies provides insights into their formation and evolution since it results from a complex interplay 
%of processes that create and reshuffle metals, including star formation, chemical enrichment, stellar and supernovae feedback, outflows, infall of primordial gas, and merging \cite[e.g][]{Dayal13}. 
%The present-day distribution of metals across dwarf galaxies results from a complex interplay between chemical enrichment, loss of metals, gas accretion and feedback. 
The advent of integral-field-units and multi-object spectrographs during the past decade or so has permitted the spatial characterization of element abundances within dwarf galaxies  at a
higher level of detail than possible before with long-slit observations. This has promoted a wealth of studies specifically aimed at quantifying the degree of chemical homogeneity across these systems and at revealing the possible presence of metallicity radial trends or localized metallicity variations. 

%The multitude of studies aimed at characterizing these spatial trends have revealed a diversity of behaviors: 
%homogeneous distribution of metals throughout the galaxies, within the uncertainties  \cite{Kobulnicky96,,KobSkill97,Lee06,Cairos09,Croxall09,Lagos13,Lagos16,Kehrig20}, localized metallicity variations \cite{Kobulnicky97,James13a,James13b,kumari17,lagos18,James20}, or well-ordered metallicity gradients comparable to those observed in spiral galaxies \cite{Lee06,Papaderos06,Annibali15,Pilyugin15,Annibali17,Annibali19}. 
 
Unlike spiral galaxies, where negative metallicity gradients have been found to be prevalent \cite{Bresolin09},  
SFDs exhibit a variety of radial behaviors. 
 Most of them show, within observational uncertainties of $\sim$0.1-0.2 dex, 
a rather homogeneous spatial distribution of metals  within $\sim$1 kpc scale \cite{Lee06,Cairos09,Kehrig20}, 
suggesting a rapid dispersion and mixing of the ejecta from stellar winds and supernovae across the ISM.
%although it is possible hat the non-detection of gradients may be due in some cases to insufficient depth of the spectroscopic data \citep{Annibali15}.}
% indicating that the gas is well mixed at all times,
This behavior is in agreement with the predictions of cosmological, three-dimensional hydrodynamical simulations of dwarf galaxies with a realistic feedback treatment and the inclusion of metal diffusion \cite{Fitts17,Escala18,Graus19}, which produce typically flat gas-phase metallicity gradients, with the rare occurrence of steep gradients limited to the inner galaxy regions \cite{Mercado21}.
% a rapiddispersion and mixing of the ejecta from stellar winds and supernovae across the ISM,  

 However, well-defined and spatially extended metallicity gradients, comparable to those observed in spiral galaxies,  have been observed in a few systems \citep{Papaderos06,Annibali15,Pilyugin15,Annibali17,Annibali19b}
 % (NGC~6822 \cite{Lee06}), Papaderos06, NGC~1705 \cite{Annibali15,},NGC~4449 \cite{Annibali17}, 
%DDo~68 \cite{Annibali19} and several dwarfs in the sample of \cite{Pilyugin15}), 
and appear to be correlated with the presence of steep inner surface brightness profiles resulting from enhanced central star formation \cite{Pilyugin15}. The reason for the presence of gradients in a few dwarfs and the absence in the majority of them is still not well understood, and it is possible that well-defined trends are hidden under  insufficient depth of the spectroscopic data in some cases \citep[e.g.][]{Annibali15}. 
Also on the theoretical aspect, spatial metallicity properties resulting from hydrodynamical simulations  should be interpreted with caution, since they are critically impacted by the specific feedback implementation, as well as by the inclusion of advection and turbulent metal diffusion, all processes that ultimately affect the metal mixing timescale \cite{Avillez02,Emerick19}.

Besides large-scale galaxy gradients, localized  metallicity variations have been observed in a few SFDs. Regions of oxygen deficiency are 
sometimes associated to increased star-formation rates and/or kinematical inflow signatures \citep{Almeida14,Kumari17,Lagos18}, suggesting rapid accretion of metal poor gas feeding ongoing star formation \cite{Verbeke14,Ceverino16}. One of these examples is the starburst dwarf galaxy NGC~4449, known to be in the process of merging with one or two smaller systems \cite{Martinez12}, which displays a lower central HII metallicity compared to more external regions \cite{Kumari17}. 
% (e.g., ; UM~448 \cite{James13b}; SDSSJ1647+21 and SDSSJ2238+14 \cite{Almeida14}; NGC~4449 \cite{kumari17};  Mrk 600 \cite{lagos18} ;  JKB~18 \cite{James20}, suggesting rapid accretion of metal poor gas feeding ongoing star formation \cite{Verbeke14,Ceverino16} .

A different scenario applies to cases of enhanced nitrogen and/or helium abundance in the proximity of Wolf Rayet (WR) stars \citep{Walsh89,Pustilnik04,James09,Lopez11,West13},
where the anomalous element ratios are likely caused by enriched ejecta from WR winds. A well-studied example is the 
SFD NGC~5253 which contains a large number of extremely young, compact stellar clusters, and it is famous for displaying a factor 2-3 nitrogen enhancement \cite{Walsh89,Kobulnicky97} in its central regions as well as evidence of N-rich and C-rich WR stars \citep{Schaerer97}. 
%(e.g, in NGC~5253 \cite{Walsh89,Kobulnicky97,Schaerer97,Lopez07,Monreal-Ibero10,West13}, Mrk~996 \cite{Thuan96,James09,Telles14}, NGC~4670 \cite{Kumari18}, HS~0837+4717 \cite{Pustilnik04}, IC~10 \cite{Lopez11}, Haro~11 \cite{James13a}; Mrk~178 \cite{Kehrig13}, IIZw70 \cite{Kehrig08}) and have suggested to be caused by enriched ejecta from WR winds. %This is supported by the results of \cite{Brinchmann08} who, using a large sample, showed that WR galaxies show an elevated N/O relative to non-WR galaxies.  
However, the association between N or He enrichment and WR stars is far from clear-cut, since cases exist where 
%WR stars have not been detected in correspondence of enhanced N/O ratios XX), or vice-versa 
no nitrogen or helium enhancement has been observed at the location of WR stars 
\citep{Kobulnicky96,Annibali17,Kehrig16}, or where, vice-versa, no WR stars have been detected 
in correspondence of enhanced N/O ratios \cite{Perez11}.
%(e.g., in NGC~4214 \cite{Kobulnicky96}, NGC~4449 \cite{Annibali17}, IC~3521, CGCG~038-051, CGCG~041-023 and SBS~1222+614 \cite{Paswan19}, IZw18 \cite{Kehrig16}), or where, vice-versa, no WR stars have been detected 
%in correspondence of enhanced N/O ratios \cite{Perez11}.  
It is possible that nitrogen-rich ejecta are still in a hot (T$\sim$10$^6$ k) phase at early WR evolutionary stages ($\lesssim$5 Myr) and will cool down and mix into the optically observed ionized
gas only after a completed WR phase, about $\sim$7-8 Myr from the onset of the starburst \cite{Monreal12}.
Furthermore, processes other than self-enrichment by massive stars have been suggested to affect the N/O spatial distribution within dwarf galaxies, such as delayed nitrogen production by intermediate-mass stars during a period of relative star formation quiescence when oxygen is no more released \cite{Garnett90} or even rapid infall of metal-poor gas which strongly dilutes the oxygen abundance \cite{Koppen05,Luo21}.

%Cosmological hydrodynamical simulations predict that stellar metallicity gradients are common in dwarf galaxies, and result from the competing effects of i) progressive radial migration of old, metal-poor stars from the center to the outskirts due to 
%feedback-driven potential fluctuations, which results in metal richer and younger populations in the central regions, and ii) late-time star formation fueled by the accretion of extended, recycled (enriched) gas, which tends to weaken the gradients \cite{Graus19,Mercado21}. On the other hand, gas-phase metallicity gradients are predicted to be weak 

%The MZR results
%from the interplay between gas accretion and recycling, star formation and feedback-driven outflows (e.g. Edmunds 1990; Dave,ÂŽ
%Finlator & Oppenheimer 2012; Feldmann 2013; Lilly et al. 2013;
%Lu, Blanc & Benson 2015), so it is widely used to constrain feedback models in cosmological simulations and semi-analytic models
%of galaxy formation (e.g. Dave, Finlator & Oppenheimer ÂŽ 2011; Lu
%et al. 2014; Torrey et al. 2014; Ma et al. 2016b).

\subsection{Chemical evolution models} \label{sec_chemev}

SFDs are gas-rich and metal-poor, some of them extremely metal-poor, and all those where faint old stars could be reached by the available photometry turned out to have formed stars over their entire life. The obvious question then is: if a galaxy has formed stars for 13 Gyr, even if at a relatively low rate, how did it manage to remain gas-rich and metal-poor? Is it because it did not produce the metals (invoking {\it ad hoc} non-standard nucleosynthesis or an initial mass function with few massive stars), or because its produced metals were diluted by large amounts of very low-metallicity accreted gas (infall), or because it got rid of the elements produced by its stars ejecting them out (winds)?  These three possible solutions were already envisaged by \cite{mc83} almost 40 years ago. Since then significant improvements have been introduced in chemical evolution modelling: direct derivation of the SFH from the CMDs of the resolved stars, observational data on gas flows, more accurate stellar nucleosynthesis, more sophisticated treatment of the dynamics involving the systems. However, a consensus on the solution has not been reached yet. 

One of the first pioneering papers on the chemical evolution of SFDs \cite{Lequeux79a} explained the low oxygen abundance measured in their HII regions invoking an initial mass function with few massive stars, thus significantly reducing the oxygen production. This early suggestion is consistent with the recent argument \cite{igimf} that a galaxy integrated initial mass function depends on the SF rate and allows for only few massive stars when the rate is lower than 1 M$_{\odot}$/yr. The galaxy integrated  initial mass function may then explain the low abundances in XMPs that have low rate, as Leo P and Leoncino. However, it is not sufficient to explain the low oxygen in XMPs with high SF rate, like IZw18.

Infall of  gas seems always necessary to replenish the gas consumed by star formation, but it is not always sufficient to keep the metallicities as low as observed. Some authors suggest \cite{gavilan13} that it could suffice if the infalling gas is primordial (i.e., metal-free). However, recent hydrodynamical models of DDO~68 \cite{Pascale21} show that, even if primordial, the infall mass necessary to sufficiently dilute the produced oxygen is inconsistent with any meaningful dynamical solution, unless the measured metallicity refers to the accreted material and not to that of the main galaxy, as mentioned in Section 3.1.1.

More successful chemical evolution models of SFDs have invoked both infall and winds to reproduce the observed low chemical abundances \citep[e.g., ][]{mt85,Pilyugin92,marconi94,romano06,romano13}, suggesting that the outflows triggered by supernova explosions in these shallow potential well systems may be very efficient in ejecting some of the elements (like oxygen) out of the system. Hydrodynamical simulations of dwarf galaxies \cite{derco99,MacLow99,scan10,robles17,romano19} have shown indeed that the winds blowing out of these galaxies can remove substantial amounts of the elements produced by the supernovae themselves, without removing much of the overall gas content; on the other hand, feedback from active galactic nuclei (AGN) has started to be included in models of dwarf galaxy evolution since only very recently \citep[e.g.,][]{Koudmani21}.
Galactic winds are difficult to detect, and only a handful of SFDs are known to host them (e.g. \cite{mcquinn19} and references therein).  Nowadays, supernova feedback appears as a viable scenario for the evolution of SFDs with fairly high SF rate, but it may not be as efficient in dwarfs with low SF, where few supernovae are expected to explode, and may not be sufficient to trigger the winds. In these cases, the galaxy integrated  initial mass function may be more effective. 

Further studies, taking into account all the involved mechanisms, appropriate stellar nucleosynthesis, possible galaxy integrated initial mass function and hydrodynamics of the galaxy system, and aimed at reproducing all the observed properties (SFH, chemical abundances, gas and star distribution and kinematics) are clearly necessary.

\section{Star Forming Dwarfs  in a cosmological context}

SFDs play a key role in cosmology and galaxy evolution. As mentioned in the Introduction, the $\Lambda$CDM paradigm predicts that the smallest dark matter haloes are the first to collapse, and are the building blocks of larger structures following a continuous hierarchical merging process \citep{White78}. 
%Since dwarfs are expected to host the first stars formed in the Universe \citep{Dayal18} they are the best candidates for Reionization at z$>$6 \citep{Bouwens12,Wise14,Trebitsch18}, and play a key role in the enrichment of the intergalactic medium (IGM) \citep{MacLow99,Scannapieco02,Ricotti08}.
%It is now well established that the IGM contains the vast majority of the baryons in the Universe
%\citep{Cen06,Shull12,Peeples14,Nicastro18} and that it was enriched with metals far from sites of star formation already at z$\sim$ 5 \citep{Pettini03}.
%Unfortunately, detailed studies of high redshift dwarf galaxies are severely challenged by the insufficient sensitivity and spatial resolution of the spectro-photometric data \citep{Oesch13,Oesch18,Bhatawdekar19}.
%Nearby SFDs, with star formation, gas content, and metallicity properties commensurate to those of high redshift dwarfs, are valuable laboratories for detailed studies of the physical processes that characterized the evolution of primeval systems in the early Universe. 
In the following, we discuss some aspects related to the study of nearby SFDs and relevant for cosmology.  

%Nearby SFDs are known to often host young super star clusters (e.g., âO'Connell et al.1994; Massey, & Hunter 1998; Boker et al 2001â), the best local analogs of the high-z, massive, young clusters detected in recent years at high redshift, which may have contributed to re-ionize the IGM (e.g., âVanzella et al 2019a,bâ). Also, galactic outflows enriched in metals are detected in a significant number of local late-type dwarfs (âMeurer et al.1992; âMarlowe et al. 1995; Della Ceca et al. 1997; Martin et al. 1999; Heckman et al. 2001; âOtt et al. 2005; McCormick et al. 2018â).

\subsection{$\Lambda$CDM and hierarchical merging at the dwarf scales} 

Although the $\Lambda$CDM paradigm has proven remarkably successful at explaining the origin and evolution of large scale  structures in the Universe, a number of potential challenges have been identified when confronting observations with predictions for dwarf galaxies \cite{Bullock17}. While some of the problems can be solved through a proper modelling of baryonic physics, alternative dark matter models offer solutions to the identified discrepancies too \citep{Vogelsberger20}. Studies of the satellite population around nearby dwarf galaxies provide a valuable testing ground for the different cosmological models. 

$\Lambda$CDM numerical simulations predict the same relative amount of substructure within dark matter haloes of different masses \cite{Diemand08} and thus present-day dwarf galaxies are expected to be surrounded and orbited by a large population of satellites and stellar streams similar to what is observed around giant spiral and elliptical galaxies \citep[e.g., ][]{Ibata01,Belokurov06,Delgado10,Crn15}. 
Unfortunately, studies of satellites around dwarf hosts with M$_{\star}\lesssim10^9$ M$_{\odot}$ have received little attention so far, mostly because of the difficulty in detecting very low-surface brightness companions or merger signatures around dwarf systems. These studies are further complicated by the steep decrease toward lower masses of the stellar-to-halo-mass ratio \cite{Sawala15,Read17}, suggesting that a large number of dwarf satellites are expected to be extremely faint or even totally dark. 
Hydrodynamical simulations predict that the merger of a purely dark-matter satellite with a gas-rich dwarf may leave morphological and kinematical peculiarities in the host stellar and gaseous components as well as trigger starbursts at pericenter passages {\cite{Helmi12,Starke16a,Starke16b}, providing indications on how to unveil the presence of interacting satellites. 
%Hydrodynamical simulations provide indications on how reveal those events, 
%showing that the merger of a  dark satellite with a gas-rich dwarf should leave morphological and kinematical peculiarities in the stellar and gaseous components, and trigger starbursts at pericenter passages {\cite{Starke16a,Starke16b}. 
%h as  imprinted on the host, as well as by star formation triggered at pericenter passages {\cite{Starke16a}. 
%Starbursting dwarfs with asymmetric distributions and velocity offsets between the gaseous and stellar components are promising candidates to unveil the presence of interacting dark satellites \cite{Starke16b}.

% Figure 5 
% Figure 5  
\begin{figure}
\centering
\includegraphics[width=\textwidth]{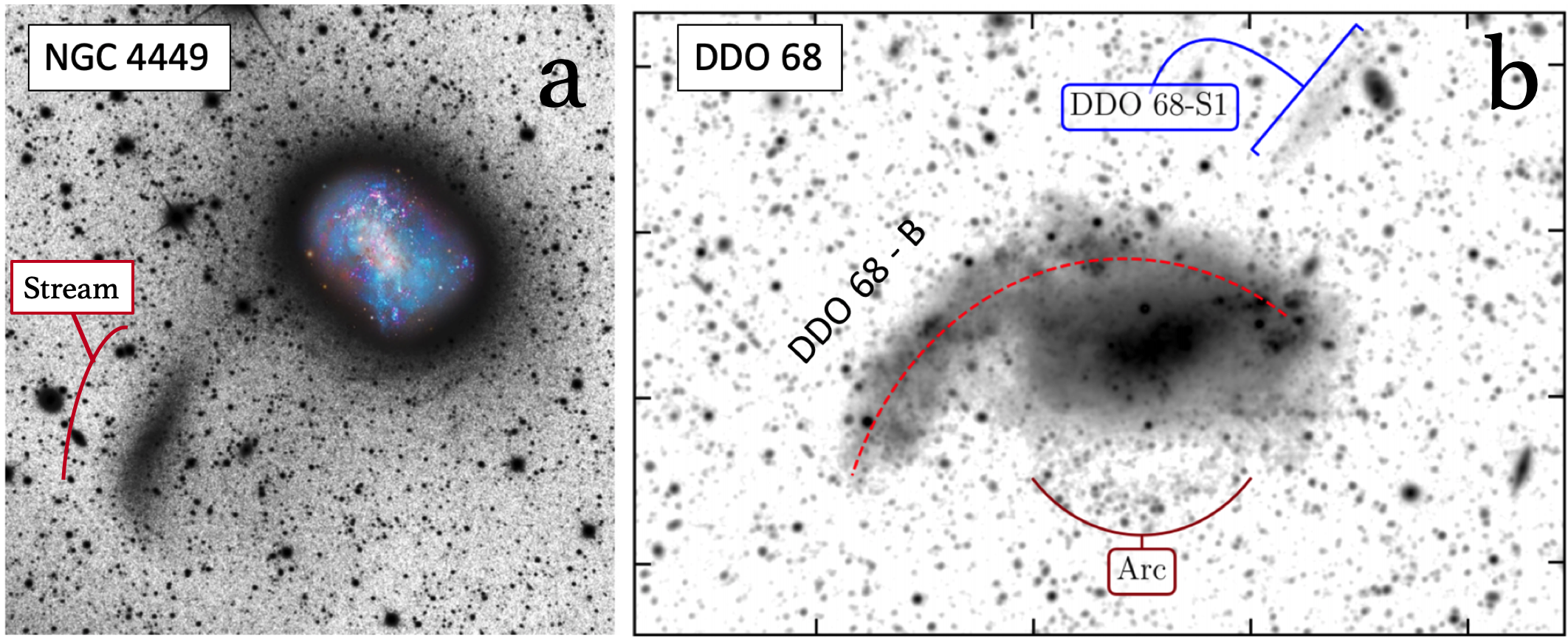}
\caption{ Examples of discovered dwarf-dwarf mergers. Panel a: from \citep{Martinez12}, the starburst dwarf NGC~4449 and its low surface brightness halo stream.  Panel b: deep LBT images of  DDO~68 caught in the process of accreting a ten times less massive companion (DDO~68 B) and an even smaller satellite (DDO~68 S1), figure from \citep{Pascale21}.}
\label{merging}
\end{figure}

So far, observational evidence for dwarf-dwarf mergers is limited to a few circumstantial cases \citep{Rich12,Martinez12,Belokurov16,Amorisco14,Annibali16,Zhang20}. Striking examples (see Fig.~\ref{merging}) are the starburst galaxy NGC~4449 \citep{Rich12,Martinez12}, associated with a low surface brightness stellar stream, the XMP void dwarf DDO~68, which is merging with at least two smaller systems ~\citep{Cannon14,Annibali16,Annibali19a}, and the dwarf spheroidal And~II, where the presence of  a kinematical stellar stream indicates a past merger between two smaller systems \citep{Amorisco14}. 
However, only in recent years significant effort has been invested into systematic searches of faint 
companions around large samples of dwarf galaxies through deep ground-based surveys with wide-field imagers on 4-10 m telescopes. 
The Solo survey \cite{Higgs16}, the MADCASH survey\cite{Carlin16}, and the SSH survey \cite{Annibali20} are scrutinizing the outskirts of local dwarf galaxies based on a resolved star approach, while other studies focus on more distant samples from large-area, deep-sky surveys to characterize  the dwarf galaxy demographics on a larger statistical base  \cite{Stierwalt15,Paudel18,Kado20}. 
These studies show that paired-dwarfs exhibit, when compared to non-paired analogues, enhanced star formation 
{\cite{Stierwalt15,Pearson16}, and that the presence of tidal features correlates with the star formation activity
\cite{Lelli14,Kado20}, demonstrating the crucial role played by mergers and interactions in the evolution of such fragile systems.  Following-up on the study of \cite{Lelli14b}, \cite{Mcquinn15b} demonstrated 
that the starburst activity is directly connected to the inner shape of the
galaxy's potential well, which may change through interactions.

\subsection{SFDs and the reionization of the Universe}

Cosmic reionization is one of the most important events in the history of our Universe, reflecting the formation of the first luminous structures on the one hand, and affecting the subsequent formation and evolution of galaxies on the other \citep{Dayal18}. Theoretical \citep{Chou07,Dayal20} and observational \citep[e.g., ][]{Finkel12,Bouwens12,McLeod16} studies suggest that dwarf galaxies may have provided the bulk of the ionizing radiation at high redshift, both because they are the most abundant galaxy population at any cosmic time \cite{Bullock00,Bouwens15} and because they are good candidates for ``leaking'' a sufficient fraction of ionizing photons ($f_{esc}$) through low density holes carved in the ISM by efficient stellar feedback in their shallow potential well \citep{Choi20}. 
%In fact, thanks to their shallow potential wells, dwarf galaxies are vulnerable to feedback processes that carve low density holes in the ISM, facilitating the loss of ionizing photons, gas, and metals via galactic winds. 
%However, confirming this scenario requires to constrain the fraction of the ionizing photons that emerge out of the dwarf galaxy environment to asses whether this is sufficient to ionize the IGM. 

While direct detection of leaking ionizing photons at z$>$4 is almost impossible due to strong absorption 
by the neutral intergalactic medium
\citep{Prochaska10}, SFDs at low redshift have been widely used as local laboratories to constrain $f_{esc}$
%\citep{Leitherer95,Bergvall06,Heckman11,Leitet13,Borthakur14,Izotov16a,Izotov16b,Leitherer16,Chisholm18}. 
\citep{Leitherer95,Bergvall06,Heckman11,Izotov16a,Leitherer16}. 
These studies show that extreme feedback powered by massive, compact star forming regions at the center of the most active blue compact dwarfs favor the escape of ionizing photons. These systems typically exhibit large [O~III]$\lambda$5007/[O~II]$\lambda$3727 ratios (from which the name of ``Green Pea'' galaxies), likely as a consequence of feedback-generated cavities in the ISM which produce conditions compatible with those of an optically thin medium \citep{Nakajima14,Vanzella16}. An example is the low mass (M$_{\star}\sim10^9$ M$_{\odot}$), low metallicity ($\lesssim$0.2 solar), Green Pea J0925+1403 at z$\sim$0.3, for which Lyman continuum emission was directly detected with 
HST COS and an escape fraction of $f_{esc}\sim8\%$, sufficient to ionize a $\sim$40 times more massive intergalactic material, derived \cite{Izotov16a}. In the more nearby Universe, at a distance of $\sim$3 Mpc,  
the first  $f_{esc}$ spatial map of the  starburst dwarf NGC~4214 was obtained by \cite{Choi20} through a novel method based on the comparison between the ionizing radiation from massive stars with the amount of photons absorbed by dust or consumed in HII regions. Most recently, \cite{Eggen21} have shown that the SFD Pox 186 is consistent 
with having blown away its entire neutral interstellar medium, which would
result in an $f_{esc}$ of nearly 100\%.

The UV background produced by SFDs leaking ionizing photons at high redshift is also expected to play a major role in suppressing star formation in the smallest dark matter haloes,
and it has been argued that ultra-faint dwarf galaxies in the Local Group may be the fossil relics of a population of dwarf galaxies formed before re-ionization \cite{Efstathiou92,Bovill09}. 
However, detailed star formation histories derived from resolved-star CMDs in the Local Group indicate a complex scenario, with some ultra-faint dwarfs having formed almost the totality of their stellar mass before z$\sim$6  ($\sim$12.8 Gyr ago)  and thus being compatible with quencing from re-ionization, but others exhibiting significant star formation also at later times 
 \cite{Brown2014,Weisz2014}. Indeed, some systems show evidence for a rather late quencing most likely due to their infall in the densest regions of the Local Group \citep[e.g., ][]{Monelli2016}. Recent cosmological hydrodynamical simulations provide some clues to this diversity of behaviours showing that, among similarly small mass haloes, those  exhibiting lower central 
 concentration are more susceptible to the effects of reionization-induced feedback to the point that may fail to form  any stars  \cite{Fitts17}. 
 
In summary, studies of the stellar populations and chemical properties in local SFDs, coupled with recent results on the  escape fraction of ionizing photons, suggest that similar systems may have been responsible for enriching with metals and re-ionizing the intergalactic medium in the early Universe. Future studies based on new generation telescopes will further shed light on these topics, as we discuss in the next section.

%\bibitem{Graus2019b}Graus, A.~S., et al. {\em How low does it go? Too few Galactic satellites with standard reionization quenching}. {\em MNRAS}, {\bf 488}, 4585-4595 (2019)

\section{Outlook}

In the 2020s our knowledge of the stellar and chemical properties of SFDs, and of galaxies in general, shall greatly improve thanks to the advent of new generation telescopes, in particular the James Webb Space Telescope (JWST), scheduled to start operations in 2022, and the Extremely Large Telescopes (ELTs),  some years later. JWST and the ELTs will operate in the near and mid-infrared with unprecedented power and resolution, thus opening new windows on ISM and stellar properties currently difficult or impossible to observe, because typical of objects that emit mostly in the infrared (e.g., late-type stars or dust grains) or  highly obscured by dust at optical wavelengths (e.g., very young stars).

JWST will allow us to resolve individual stars at much fainter magnitudes than HST, thus significantly enlarging the range of distances where extremely ancient epochs can be reached. This will lead to a significant step forward in our understanding of the earliest SF activity in dwarfs, and on their contribution to the early evolution of the Universe. One of the approved Early Release Science Programs (PI D. Weisz) that will be executed as soon as JWST becomes operational is devoted to resolve the individual stars in one SFD and two other stellar systems in the  Local Group. Its aim is to determine the best observational setups and develop data reduction techniques useful for an optimal exploitation of JWST for studies of resolved stellar populations and SFH derivations. 

The superb imaging and spectroscopic capabilities of JWST in the infrared will allow a major improvement in the combined study of ISM, feedback, and stellar populations in local proxies of high redshift re-ionization sources, and will also provide for the first time a direct view of rest-frame ultraviolet and optical emission from galaxies in the actual re-ionization era  \citep{Giri20}. 

When the ELTs will be operating, their extreme resolving power will let us measure individual stars in the most central and crowded regions of galaxies well beyond 20 Mpc, an achievement out of reach of HST and JWST. ELTs will allow both high-resolution and multi-object spectroscopy in SFDs, fostering a much deeper understanding of the chemical properties of SFDs and of their evolution.
Since the size of the field of view of the ELTs will be severely limited by the adaptive optics requirements, the synergy with JWST will be key to cover both the innermost and outer regions of the galaxies of interest. When also Euclid and the Nancy Roman Space Telescope will come into play, their much wider fields of view will add full coverage of the most external regions, providing a complete census of low surface brightness streams around dwarf (and massive) galaxies.

The SFHs and chemical properties derived from this wealth of new data will provide the observational basis for 
models that combine detailed chemical evolution treatments with detailed hydrodynamical simulations of galaxy formation and evolution in cosmological frameworks. 
This combination is currently challenged by the very different timescales typical of chemical (long) and dynamical (short) evolutions, but new generation super computers will hopefully make it possible within a few years from now.
This will lead to a breakthrough in  the understanding of galaxy evolution. 

\vskip 0.3cm \noindent {\bf Acknowledgements}

\noindent This review has been published by Nature Astronomy on January 13$^{th}$ 2022, and is part of the Collection of short and long articles on Dwarf Galaxies that the Journal is publishing since December 13$^{th}$ 2021. 
This version of the article has been accepted for publication after peer review, but is not the Version of Record and does not reflect post-acceptance improvements, or any corrections. The Version of Record is available online at: \text{https://doi.org/10.1038/s41550-021-01575-x}. Use of this Accepted Version is subject to the publisher's Accepted Manuscript terms of use
\text\small{https://www.springernature.com/gp/open-research/policies/accepted-manuscript-terms.}
We thank an anonymous referee for the very constructive and professional comments and suggestions.

{}

\end{document}